\documentclass[%
reprint,
amsmath,amssymb,
]{revtex4-2}
\usepackage{xcolor}

\usepackage{graphicx}% Include figure files
\usepackage{dcolumn}% Align table columns on decimal point
\usepackage{bm}% bold math
\usepackage[colorlinks=true,linkcolor=blue]{hyperref}% add hypertext capabilities
\usepackage[mathlines]{lineno}% Enable numbering of text and display math
\begin{document}
	
	\preprint{APS/123-QED}

	\title{Emergent anisotropy in magnetic metamaterials}
	
	\author{Nanny Strandqvist}
	\author{Björn Erik Skovdal}
	\author{Merlin Pohlit}
	\author{Henry Stopfel}
	\author{Vassilios Kapaklis}
	\author{Björgvin Hjörvarsson}
	\affiliation{Department of Physics and Astronomy, Uppsala University, Box 516, SE-75120 Uppsala, Sweden}
	
\begin{abstract}
		
		We demonstrate directional dependence of the self-modification of internal mesospin textures in magnetic metamaterials, arising from the coupling of the atomic- and mesoscopic length-scales.
		Dressing the mesospins in different directions enables the quantification of the degree of texture in the internal magnetization and its impact on the interaction energy of the mesospins. 
		The emerging anisotropy is manifested in a directional dependence of the remanent magnetization with temperature.
		
\end{abstract}

	\maketitle
	
\section{\label{sec:introduction}Introduction}
	
    Magnetic metamaterials can be used for investigations of phase transitions, with a control of system parameters only surpassed by numerical methods \cite{kapaklis_melting_2012,levis2013thermal,Arnalds2014,arnalds2016new,Streubel2018,Leo2018,sendetskyi_continuous_2019, ASI_Review_2020}. 
    In these studies, the magnetic islands - \textit{mesospins} - have been treated as indivisible building blocks without inner structure, with a classical order-disorder transition of the mesoscopic elements. 
    It is only recently their inner degrees of freedom have been included in the description \cite{Skovda2021,Gliga_PRB_2015,sloetjes2021effect}. For disk-shaped mesospins, a collection of works addressed dynamics involving bistability of magnetization textures within, depending on geometry and temperature \cite{Ding_2005PRL, erik2014}. It was further shown that the size and shape of the mesospins determines also to which extent static internal magnetic textures, such as S-, C-, O- and vortex states, prevail \cite{shinjo_magnetic_2000,Klaui_vortx_2003,Ha2003, Ha2003_1}. A more detailed understanding of the role of thermal excitations in the magnetic texture was only recently developed \cite{Skovda2021, sloetjes2021effect}. In the light of this, systems allowing for the study of the effect of internal mesospin degrees of freedom on the collective order and thermal dynamics, have been shown to exhibit exotic order-disorder transitions \cite{Skovda2021, tricritical}. In all of these cases, circular magnetic islands effectively act as XY spins with variable spin lengths. 
	
   The emergence of an anisotropy in magnetic metamaterials has previously been demonstrated and attributed to geometrical parameters: mesospin shape and lattice symmetry \cite{Sloetjes2017, Digernes_2020cf, Digernes_APL}. Here, we show that the internal magnetic textures of disk-shaped mesospins, along with their local magnetization configurations also play an essential role. The directional dependence is shown to originate from an interplay between the extension of the elements and the symmetry of the lattice, giving rise to a self induced-anisotropy for the inter-island interactions. Thus, the internal degrees of freedom lead to an emergent magnetic anisotropy on the mesoscale, which is neither present in the parent material nor in the individual mesospins. 
    Furthermore, we show that the strength of the anisotropy can be altered by the size of the mesospin, effectively allowing for the choice of fourfold or eightfold rotational degeneracy of metastable states with collinear magnetization components.

	\section{\label{sec:method}Materials and methods}
	
    The samples were produced by post-patterning a polycrystalline three-layer thin film using electron beam lithography. 
    The thin films consisted of Pd (40 nm)  - Fe$_{13}$Pd$_{87}$ (10 nm) - Pd (2 nm) and were deposited by dc magnetron sputtering on a fused silica substrate in an ultrahigh-vacuum (base pressure below $2 \times 10^{-7}$~Pa). 
    A detailed description of the complete process can be found in \citet{Skovda2021}. 
    The periodic square arrays of circular islands were fabricated with a fixed edge-to-edge distance of 20 nm and diameters $D = $ 250, 350, and 450 nm. Each array contains 1 – 34 $\times 10^7 $ mesospins.
	
    The temperature dependence of the magnetization was determined using standard magneto-optical Kerr effect (MOKE) in a longitudinal mode \cite{Skovda2021}. 
    The incident wavelength of the $p$-polarized incident laser was set to 659 nm with a spot size of 1 mm$^2$, giving the response of approximately million discs. 
    An external field with an amplitude of 40  mT was applied along the [10] and [11] directions, with a sweep rate of 0.11 Hz. 
	
    The interplay of coupling and inner magnetization texture states, was explored by performing micromagnetic simulations using MuMax$^{3}$ \cite{Vansteenkiste2011}. 
    Disks, mimicking the actual mesospincs, were placed on a 5$\times$5 square lattice with periodic boundary conditions in lateral directions (PBC$_{\mathrm{x}}$ = PBC$_{\mathrm{y}}$ = 5, PBC$_{\mathrm{z}}$ = 0). 
    A saturation magnetization of $M_{\mathrm{sat}} = 3.5 \cdot 10^{5} $ A/M, and exchange stiffness of $A_{\mathrm{ex}} = 3.36 \cdot 10^{-12}$ J/m, were chosen based on previous work \cite{erik2014, Ciuciulkaite2019}, using the same alloys. 
    The in-plane cell size was defined as 0.5$\cdot l_{\mathrm{ex}}$, were the exchange length, $l_{\mathrm{ex}}$ is a material parameter defined by $M_{\mathrm{sat}}$ and $A_{\mathrm{ex}}$ \cite{Vansteenkiste2014}. 
    The average magnetization $|\langle \mathbf{m} \rangle |$ of the magnetic states in Table \ref{table1} were obtained by saturating disks placed on a 5$\times$5 grid along [10] and [11] direction. This was followed by relaxing the systems to its minimum free energy, leading to a mixture of different states.
    Thereafter, each island was extracted, and the net moment along one axis was used as a base for calculating $|\langle \mathbf{m} \rangle |$ and is, therefore, an average of several disks.

\section{\label{sec:results}Results and discussion}	
	
\subsection{Thermally induced transition}

	\begin{figure}[t]
		\includegraphics[width=8.25cm]{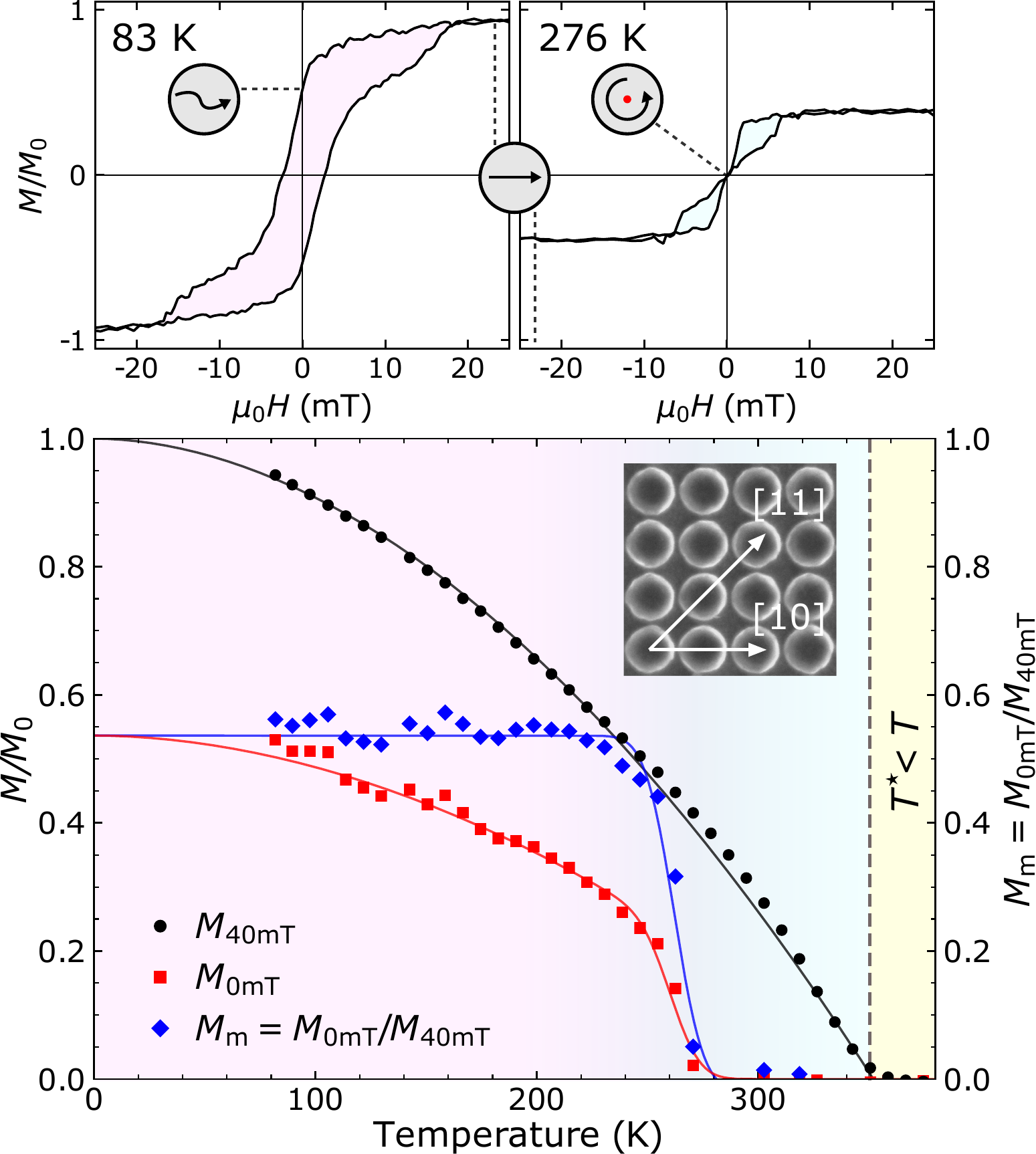}
		\caption{\label{fig:fig1} 
		Top: Magnetization loops at 83 K and 276 K for mesospins with a diameter of 350 nm and magnetic field applied along [11] direction. 
        The curves have been normalized to the net magnetization at 0 K, obtained by fitting the magnetization at 40 mT. 
        Illustrations represent possible internal magentization textures at 40 and 0 mT, respectively. 
        Bottom: Temperature dependence of magnetization $M_{\mathrm{40mT}}$ and $M_{\mathrm{0mT}}$ (left axis) and the ratio $M_{\mathrm{m}} = M_\mathrm{0mT}/M_\mathrm{40mT}$ (right axis). 
        The inset shows a scanning electron microscopy image with arrows indicating the principal lattice directions [10] and [11].}
	\end{figure}
	
    Representative hysteresis loops measured along [11] direction are shown in the upper half of Fig. \ref{fig:fig1} for disk-shaped mesospins with a diameter of 350 nm. 
    Applying a field of 20 mT results in a state with a net normalized magnetization of 0.94(1) at 83 K. 
    When reducing the applied field, the magnetization decreases, and in a zero field, a remanence of 0.52(1) is obtained.
    When increasing the temperature, the hysteresis loops change form. 
    At $e.g.$, 276 K, zero magnetization is observed at remanence while saturation is obtained already at 7 mT, with a net magnetization of 0.37(1). 
    Hence, the hysteresis loops obtained at elevated temperatures are representative of vortex states \cite{Cowburn1999}, while the hysteresis loops obtained at low temperatures have a clear ferromagnetic component \cite{Skovda2021}. 
    
    The hysteresis loops described above are clearly affected by the mesospin magnetic textures. For example, results obtained at 276 K are representative for a response of a magnetic vortex state, which is removed by applying a small external field (7 mT). 
    Small fields like this, do little to the thermally induced reduction of the magnetization. Consequently, the magnetization (at these modest fields) can be viewed as representing the intrinsic magnetization of the material ($M_{\mathrm{40mT}}$).
    A modified Bloch law is used to describe the temperature dependence: $M_\mathrm{40mT}(T) = M_\mathrm{40mT}(0) ( 1-(T/T^{\star})^\alpha)$, where $\alpha = $ 1.89(4) and $T^{\star} = $  352(1) K, represented by a black solid line in bottom panel of Fig. \ref{fig:fig1}.
	
	The temperature dependence of the remanent magnetization, $M_\mathrm{0mT}$, is symbolised by red squares in Fig. \ref{fig:fig1}. The remanence vanishes at 273 K, well below the ordering temperature of the material.
	We can extract the contribution from the magnetic texture to the changes in magnetization by dividing by the intrinsic magnetization of the material ($M_\mathrm{m}$=$M_\mathrm{0mT}$/$M_\mathrm{40mT}$) \cite{kapaklis_melting_2012, Skovda2021}. 
	Therefore, the reduction of the relative magnetization arising from the presence of inner magnetic textures can be written as $M_\mathrm{T} = 1 - M_\mathrm{m}$. 
	A plateau with $M_{\mathrm{m}} = 0.54(1)$ is observed at temperatures below 245 K. Consequently, almost half ($M_\mathrm{T} = $0.46(1)) of the moment is not contributing to the obtained magnetization.
	$M_{\mathrm{m}}$ vanishes in a relatively narrow temperature range, and at 273 K, the transition to vortex states is completed. Above this temperature, all the moment is in the magnetic textures giving rise to zero magnetization up to the ordering temperature of the material.

\subsection{Size dependence}
	
	\begin{figure}[b]
		\centering
		\includegraphics[width=8.25cm]{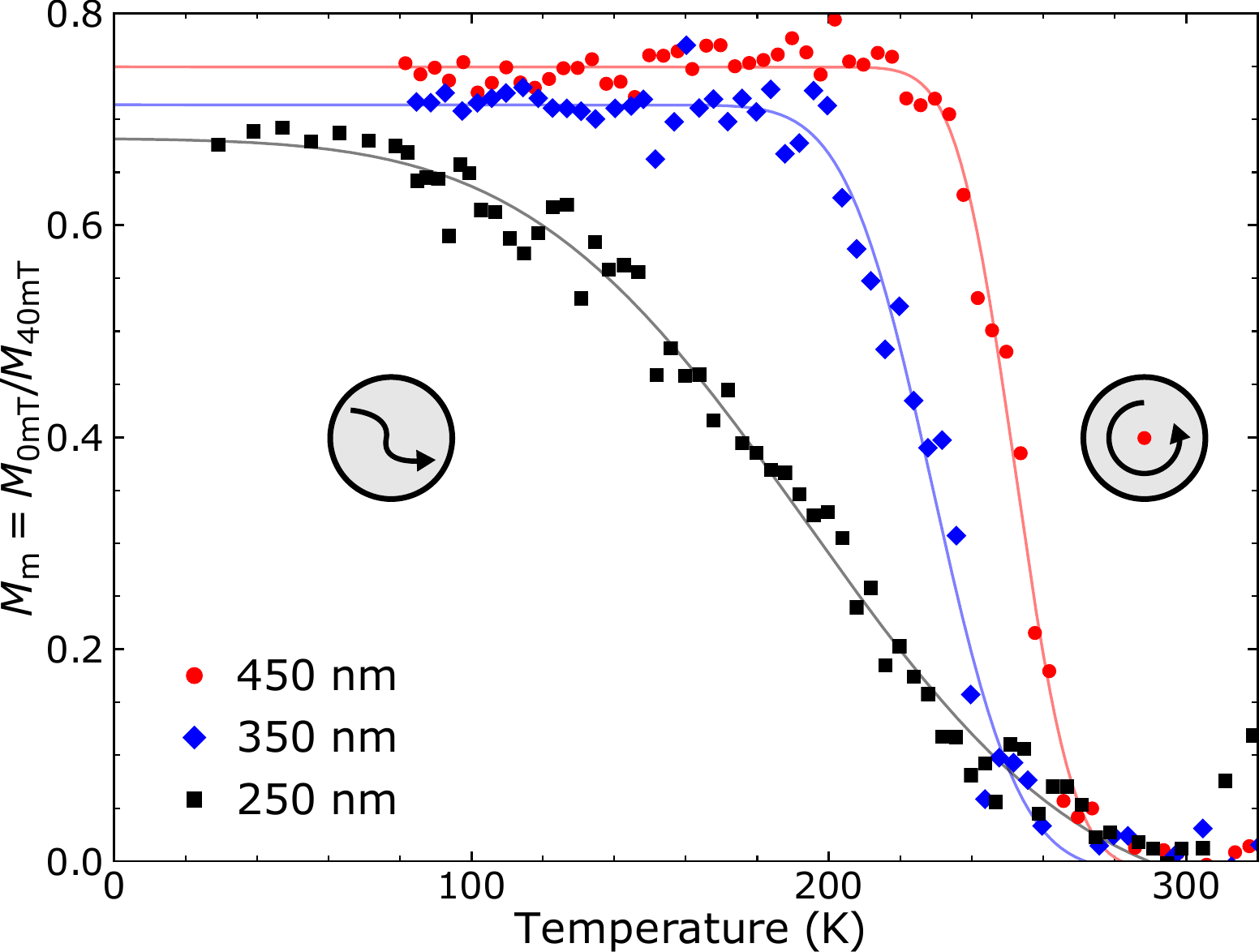}
		\caption{\label{fig:fig2} 
		Temperature dependence of $M_{\mathrm{m}} = M_{\mathrm{0mT}}/M_{\mathrm{40mT}}$ along the [10] direction for mesospins with $D = $ 250, 350 and 450 nm. 
		Error functions (solid lines) were chosen in order to determine the transition temperatures. Illustrations exemplify the possible internal texture of a mesospin.
		}
	\end{figure}

		\begin{figure}[t]
		\centering
		\includegraphics[width=8.25cm]{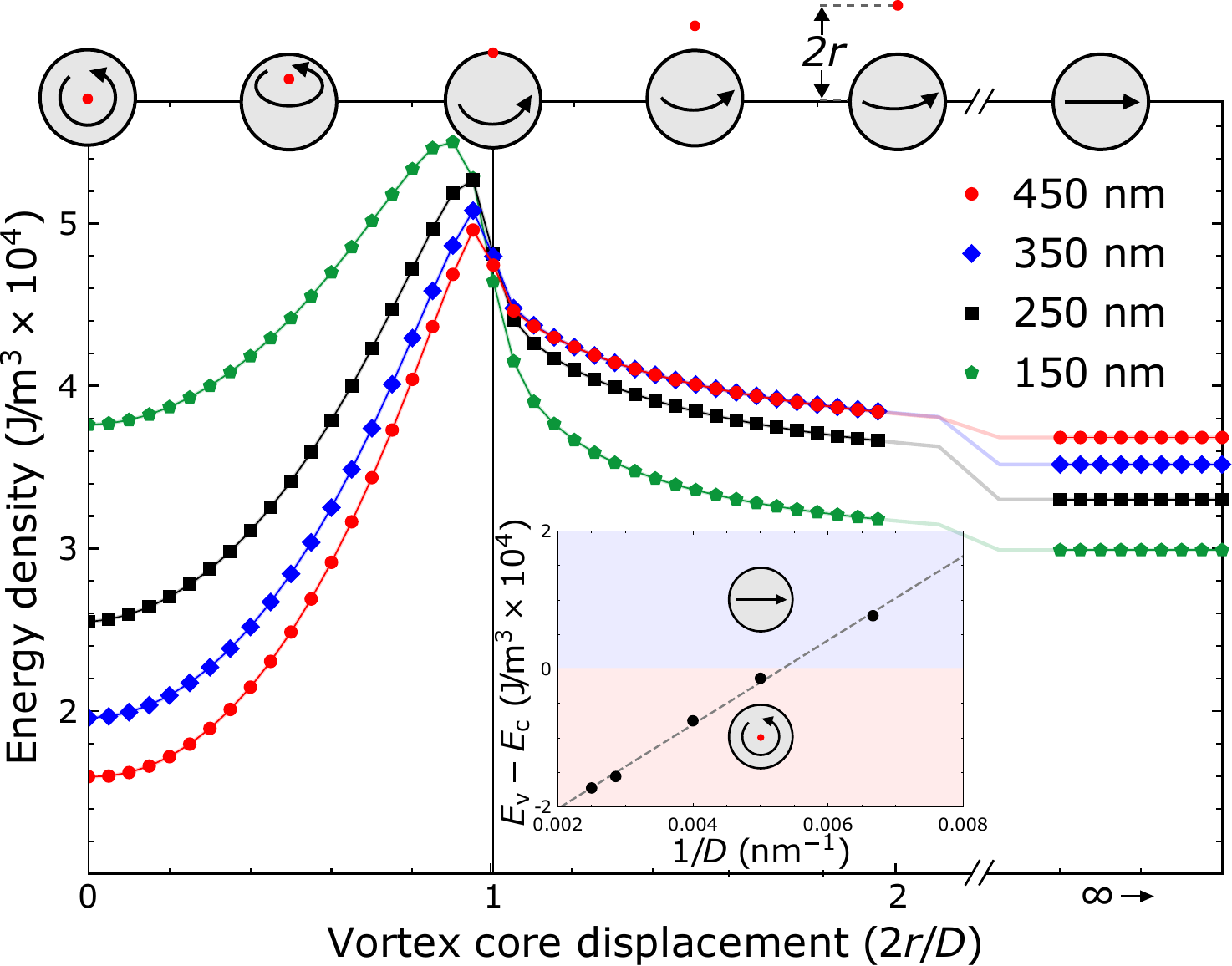}
		\caption{\label{fig:fig3}
        The free energy density of vortex, intermediate and collinear state as a function of vortex core displacement for $D = [150, 250, 350, 450]$ nm dressed along the [10] lattice direction. 
        The size dependence of $E_{\mathrm{v}} - E_{\mathrm{c}}$, display the boundary between collinear (blue shade) and vortex (red shade) states is shown in the inset. 
        The top of the figure illustrates how the core is moved with a distance of $2r$, leading to C-states with varying degree of magnetic texture until the collinear state is reached at infinity.
		}
	\end{figure}
	
	The influence of size on the transition of the mesospins is illustrated in Fig. \ref{fig:fig2},
	where the temperature dependence of $M_{\mathrm{m}}$ for mesospin diameters $D = [250, 350, 450]$ nm, with an external field applied along the [10] direction is shown. 
	All samples display a error function-like transition, with a plateau value of $M_{\mathrm{m}}$ in the range 0.68(1) to 0.75. 
	Here we note that the smaller the diameter is, the lower the temperature is for the onset of the transition. 
	No clear trends are observed with respect to the temperature when the transition is completed. 
	
	To rationalize these results, we need to get a handle on different contributions to the energy landscape, including the effect of interactions on the inner texture of the mesospins. 
	Therefore we performed micromagnetic calculations using MuMax \cite{Vansteenkiste2011}. 
	The total magnetic energy of the arrays can be defined as: $E_{\mathrm{tot}} = E_{\mathrm{t}}  + E_{\mathrm{s}} + E_{\mathrm{j}}$, where $E_{\mathrm{t}}$ is the energy cost of magnetic texture, $E_{\mathrm{s}}$, the magnetostatic energy arising from the stray field, and $E_{\mathrm{j}}$, the energy associated with magnetostatic coupling between the islands. 
	$E_{\mathrm{tot}}$ has been calculated as the vortex cores have been moved along paths, resembling one of many possible trajectories arising from the application of an external magnetic field as in the experimental procedure illustrated at the top of Fig. \ref{fig:fig3}. 
	
	From the expression described above, the energy density of the magnetic texture for $D = [150,250, 350,450]$ nm was calculated. In Fig. \ref{fig:fig3} we show the results corresponding to applying the field along the [10] direction.
	In a vortex state, the energy ($E_{\mathrm{v}}$) is only dependent on the cost of texture since $E_{\mathrm{s}} \approx E_{\mathrm{j}} \approx 0$. 
	The results reveal a size dependence, where the cost of the vortices decreases with increasing size. 
	Moreover, moving the vortex core from the center results in a collinear magnetic component with a corresponding stray field, leading to a coupling between the mesospins. 
	Consequently, the energy increases, reaching a maximum at $2r/D \approx 0.9$. 
	Continuing moving the vortex core, results in an energy decrease. When the core has been annihilated at the rim, $2r/D =1$, all mesospins possess a magnetization texture resembling a C-state. 
	The degree of texture continues to vary until the vortex cores have reached infinity and all disks are in a collinear state. Here the energy ($E_{\mathrm{c}}$) decreases with size. 
    Energy barriers separate the vortex and collinear state. 
    The height of the barrier changes with size and depends on whether it is passed from the collinear or vortex state.
    
    The inset in Fig. \ref{fig:fig3} shows the boundary between the collinear and vortex state as a function of $1/D$. 
    As the energy difference ($E_{\mathrm{v}} - E_{\mathrm{c}}$) increases with size, the linear fit reveals that for $D \lesssim$ 189 nm, the collinear state is the ground state, while a vortex state is favored when the diameter is larger.
    This is consistent with PEEM results discussed by \citet{Skovda2021}, where a mesospin diameter $>$ 189 nm was concluded to preferably have a vortex texture. 
    Interacting mesospins with $D = $ 150 nm displayed a collinear state and can therefore be viewed as two-dimensional XY-rotor.
    Thus, the vortex is the state with the lowest energy for all samples investigated experimentally in this work. 
    Consequently, the elements can be viewed as being ``dressed" by an external field, inducing a change in energy and giving rise to a ferromagnetic signal in the measurements. 
    Before addressing the importance of these two contributions, we need to investigate what happens when the direction of the net magnetization of the elements changes.
    
    \begin{figure}[b]
    	\centering
    	\includegraphics[width=8.25cm]{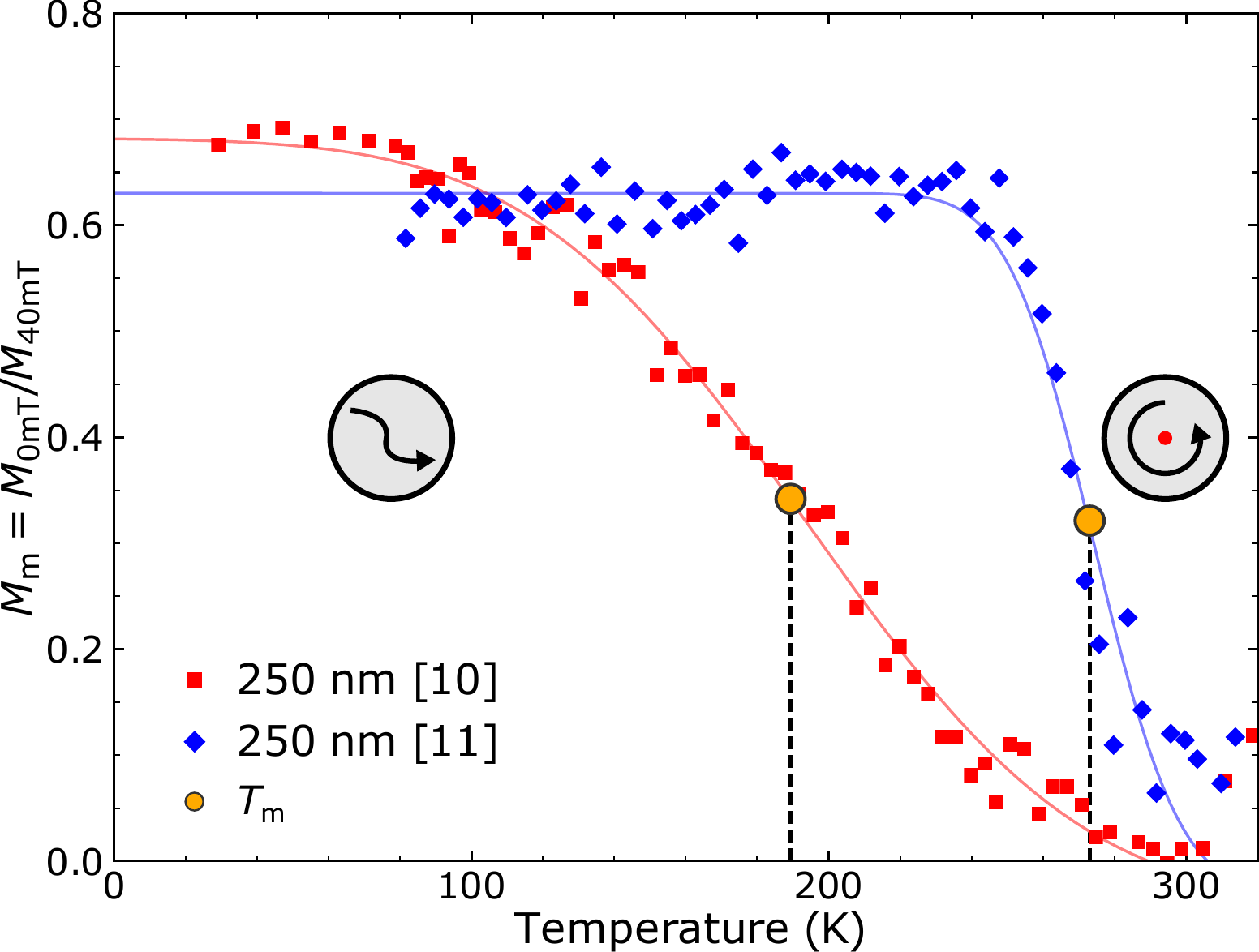}
    	\caption{\label{fig:fig4} 
    		Directional dependence of $M_{\mathrm{m}} = M_{\mathrm{0mT}}/M_{\mathrm{40mT}}$ as a function of temperature with external field applied along [10] and [11] direction for mesospins with a $D = $ 250 nm. 
    		Error functions (solid lines) were chosen in order to determine the transition without assigning any physical interpretation to it. 
    		Yellow dots represent the transition temperature $T_{\mathrm{m}}$ separating the collinear-like and vortex state, and illustrations exemplify possible internal states.
    	}
    \end{figure}

\subsection{Directional dependence}

	\begin{figure}%[b]
		\includegraphics[width=8.25cm]{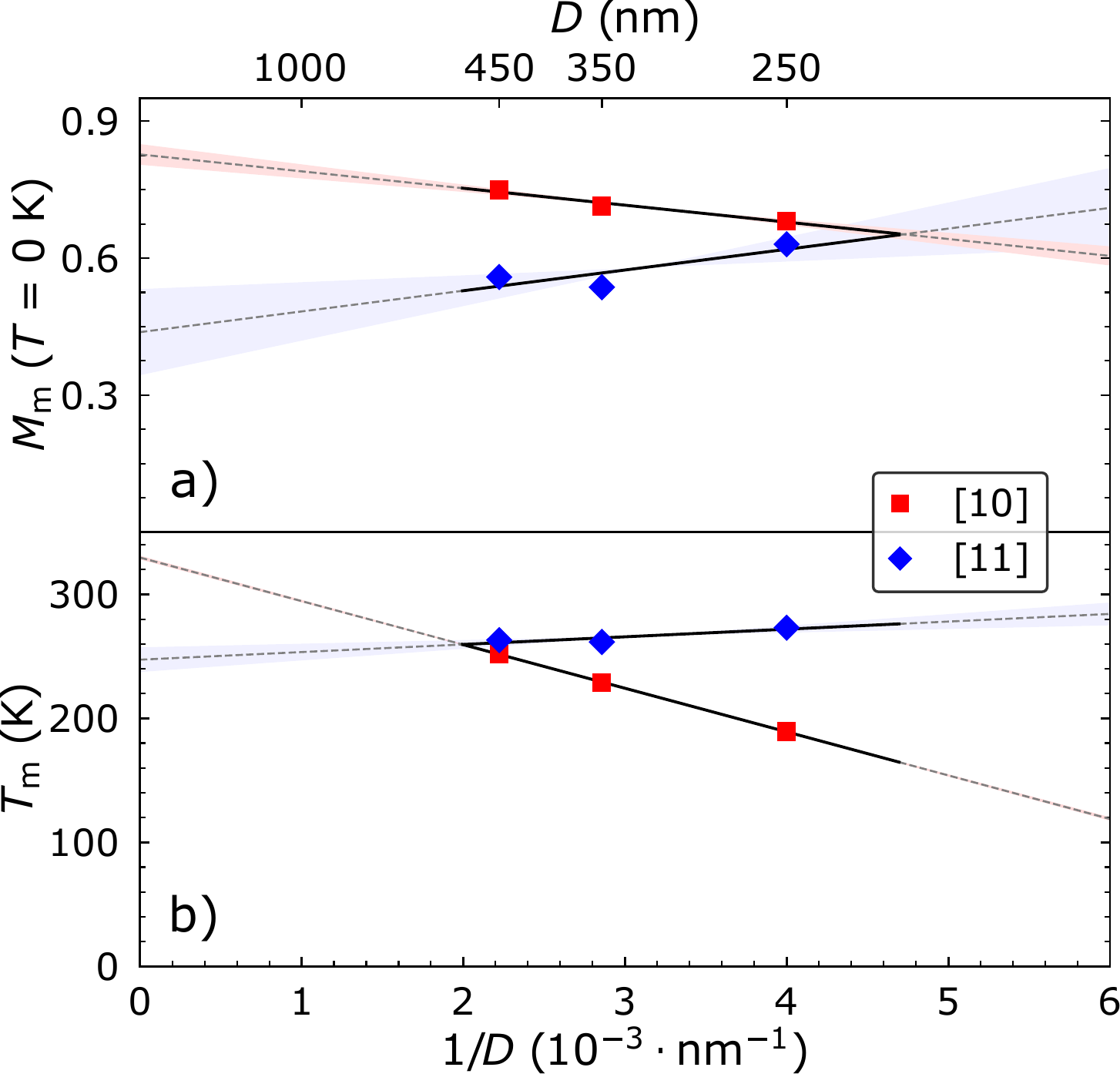}
		\caption{\label{fig:fig5} 
        Qualitative trends for a) the extent of magnetic texture, $M_{\mathrm{m}} (T = 0 \: \mathrm{K})$, and b) the transition temperature $T_{\mathrm{m}}$ separating the dressed and vortex state for $D = [250, 350, 450]$ nm along [10] and [11] direction.
        The data error bars are smaller than the symbols and are therefore not visible. 
        The shaded areas represent the estimated uncertainty for the respective data sets.
		}
	\end{figure}
	
    The directional dependence in the magnetization for mesospins with a diameter of 250 nm is displayed in Fig. \ref{fig:fig4}. A plateau value of $M_{\mathrm{m}} = $  0.68(1) is obtained when the field has been applied along the [10] direction, whilst $M_{\mathrm{m}} = $ 0.63 is obtained along the [11] direction. The lower magnetization of the array obtained when dressed along [11], is accompanied by a more stable configuration, manifested in a higher transition temperature. To allow for quantitative comparison, we define a transition temperature $T_m$ defined as $M_{\mathrm{m}}(T_m)  =\frac{1}{2}M_{\mathrm{m}} \: (T =  0$ K). By this definition, $T_{\mathrm{m}} = $ 193(3) K is obtained when the external field has been applied along [10] while $T_{\mathrm{m}} = $ 274(1) K along the [11] direction.
    
    Fig. \ref{fig:fig5} shows the effect of the chosen direction of the applied field on $M_{\mathrm{m}}$ and $T_{\mathrm{m}}$ for all sample sets. 
    The fits reveal linear relations with inverse mesospin diameter, $1/D$. 
    Positive slopes are observed when the external field has been applied along [11]. This is consistent with a decrease in magnetic texture (increasing $M_{\mathrm{T}}$) with decreasing mesospin size and an  increase in the transition temperature. 
    When the field has been applied along the [10] direction, a decrease of $M_{\mathrm{m}}$ is obtained, accompanied by a profound decrease in $T_{\mathrm{m}}$. 
    Hence, the direction of the applied field results in a significant difference in the degree of magnetic texture and in the onset of the transition. 
    The difference in response to an external field vanishes at diameters around 212 nm.
    
\begin{figure}[t]
		\centering
		\includegraphics[width=8.25cm]{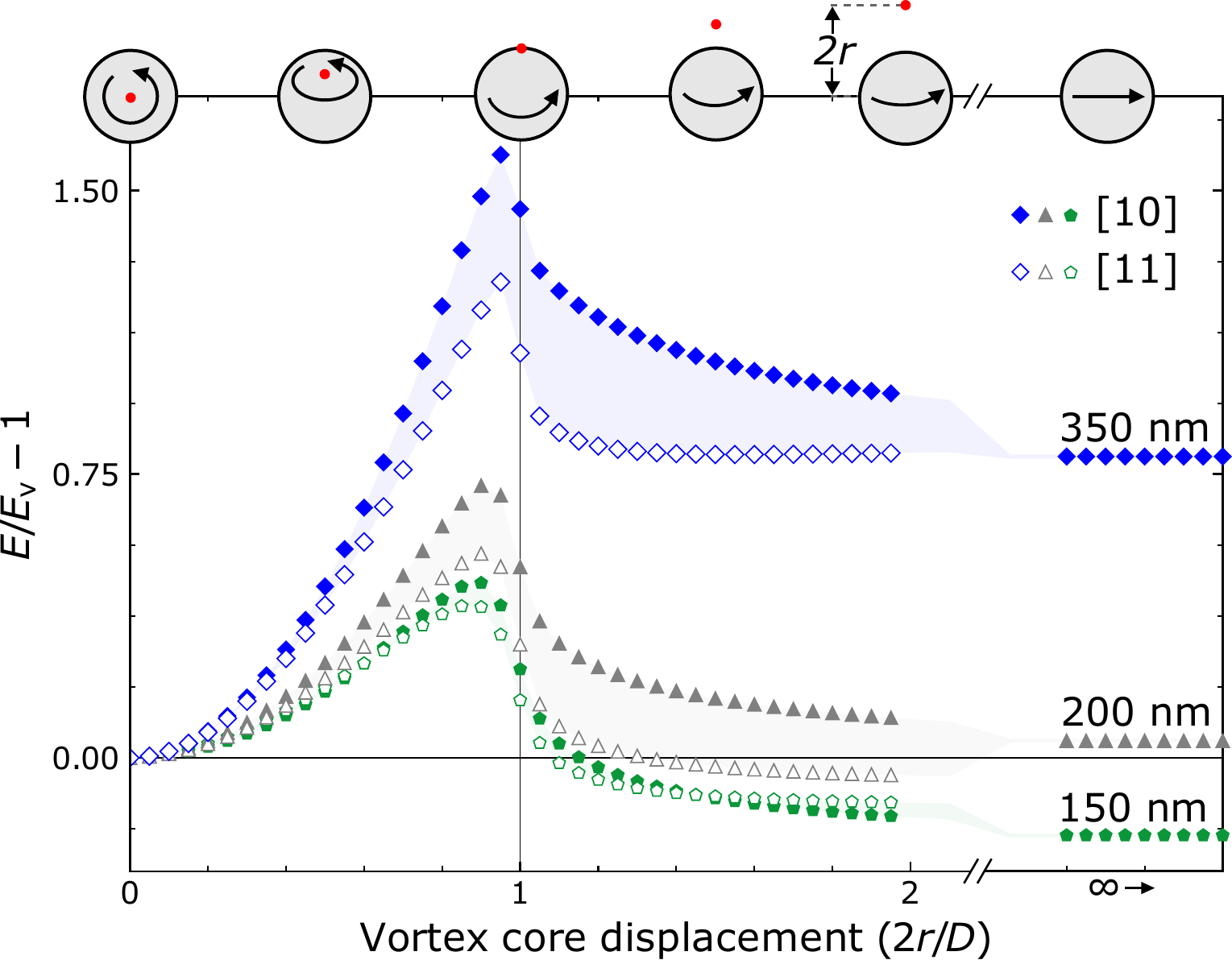}
		\caption{
		\label{fig:fig6}
        Energy landscape for the transition between vortex and collinear state for $D = [150, 200, 350]$ nm. 
        The energy has been normalized with respect to the vortex state. 
        Hence, $E/E_{\mathrm{vortex}}-1 = 0$ when the vortex core is in the middle of the disks. 
        Numerical calculations were generated by moving the vortex core along a path corresponding to [10] (empty symbols) or [11] (filled symbols) direction in an alternating manner, either up or down for every other disk. 
        At the top of the figure, illustrations show how the core moves with a distance of $2r$, leading to C-states with varying degrees until the collinear state is reached.
			}
	\end{figure}

    We now turn our attention to the impact of the direction of the magnetization on the energy landscape.
    Fig.~\ref{fig:fig6} shows the normalized energy ($E/E_{\mathrm{v}} -1$) as a function of vortex core displacement for $D = [150,200,350]$ nm dressed along the [10] (filled symbols) and the [11] direction (empty symbols).
    The ground state for mesospins with $D = $ 350 nm is the vortex state, while the collinear state is metastable and independent of orientation. 
    However, when inner magnetic textures are allowed, [10] and [11] are no longer equivalent, and the energy is observed to be lowered when the net magnetization is along the [11] direction. 
    Since the degree of magnetic texture is equivalent for the two directions, the energy gap can only be associated with an energy gain mediated by magnetostatic coupling. 
    This can be related to the experimental results observed in Fig. ~\ref{fig:fig5} where the transition temperature is higher when the external field has been applied along [11] direction.
    For $D = $ 200 nm, the lowest energy is achieved at an intermediate state along the [11] direction. 
    Implying that the energy landscape can be modified, where rotation of the net magnetization leads to an altered degeneracy of the metastable states. 
    When the diameter is 150 nm, the collinear state is energetically favored, and minor effects are observed from internal magnetic texture on the interaction of the elements. 
    
\subsection{Interplay of magnetic interactions and texture}

	\begin{table}[t]
		\begin{center}
			\caption{\label{table1} 	
				Magnetic texture and topology of V-, O-, S- and C-state depicting the nonuniformity of the mesospins. The average magnetization, $|\langle \mathbf{m} \rangle |$, is extracted along the horizontal axis of the illustrated states.
			}
			\includegraphics[width=8.25cm]{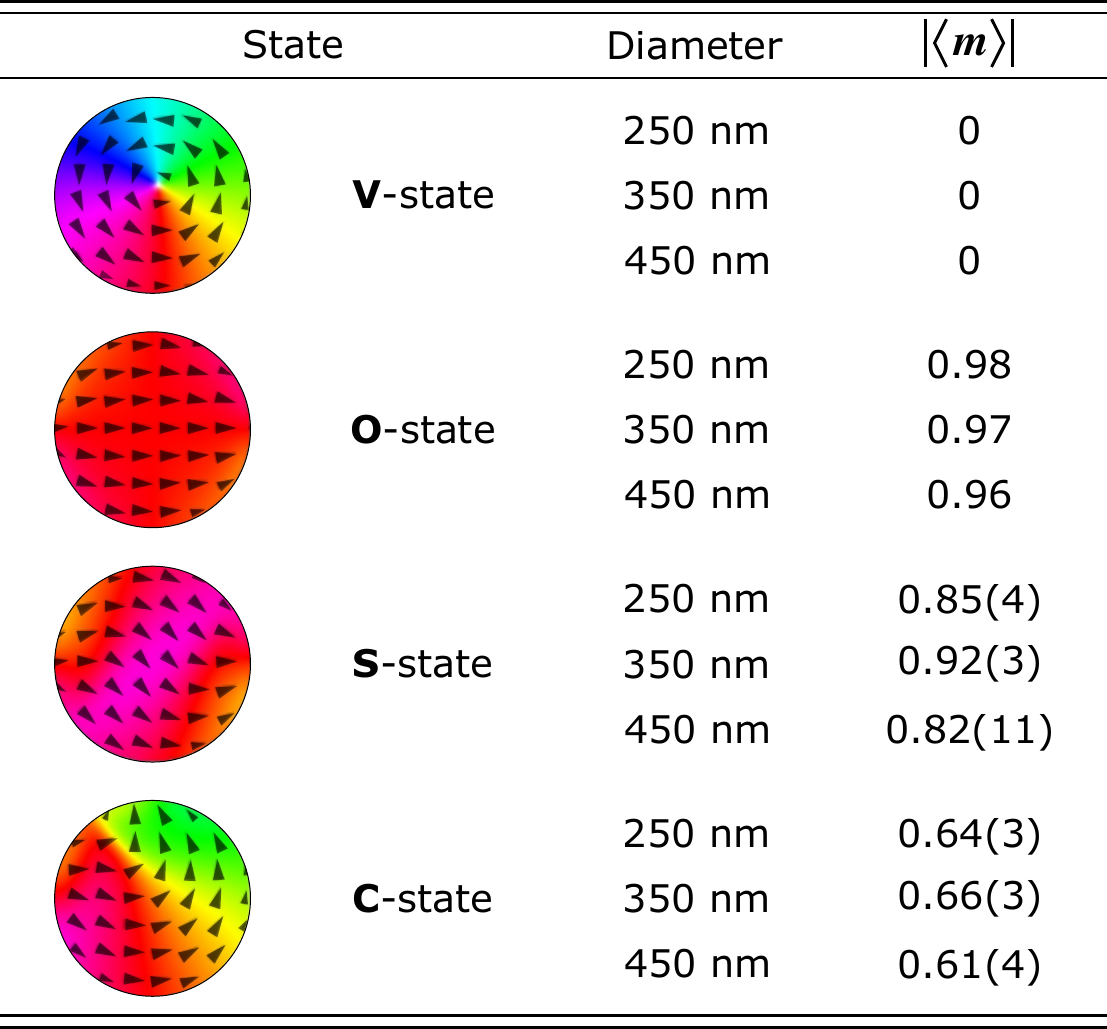}
		\end{center}
	\end{table}

    The experimental results obtained for $M_{\mathrm{m}}$ and $T_{\mathrm{m}}$ illustrated in Fig. \ref{fig:fig5}, reveal the existence of magnetic textures with transverse components and an altered onset of the transition. The extent of this is clearly dependent on the size of the mesospins and the direction of the externally applied field. To understand this, we need to take a closer look at different magnetic states and the nature of the interaction between the mesospins when the external field is applied along different directions.
    
    Even though the ground state is a vortex, the circular shape can accomodate internal magnetic textures such as O-, S-, and C- states, illustrated in Table \ref{table1} \cite{Ha2003, Ha2003_1}. The average magnetization, $|\langle \mathbf{m} \rangle |$, hereafter referred to as the mesospin moment, has been extracted along the horizontal axis of the illustrated textures shown in the table. The results from the simulations imply that the mesospin moment along the horizontal axis decrease with increasing diameter.
    
\begin{figure}[t]
		\centering
		\includegraphics[width=8.6cm]{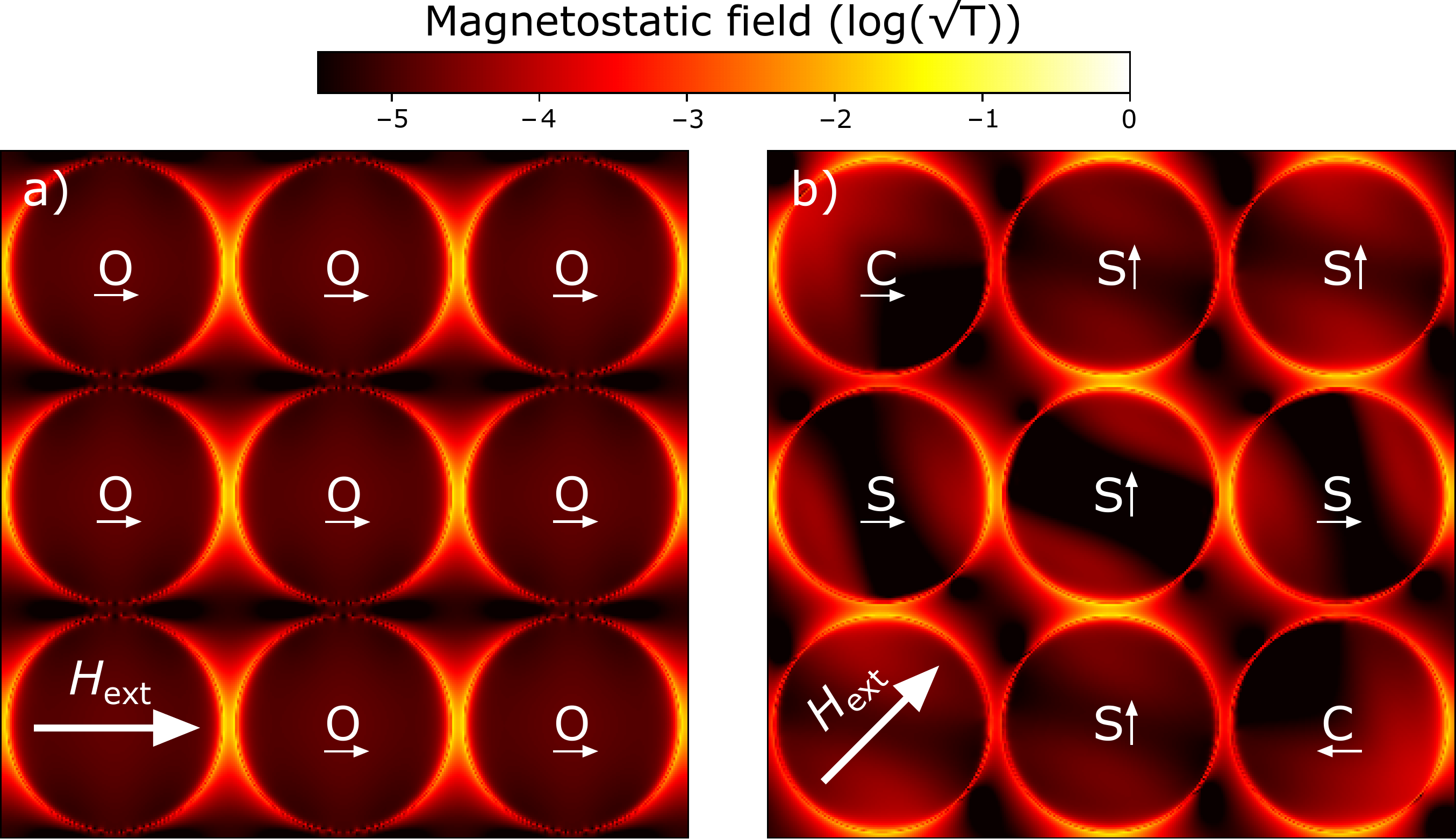}
		\caption{
			\label{fig:fig7}
			 Spatial distribution maps of demagnetizing field in relaxed state obtained from micromagnetic simulations for mesospins with a diameter of 350 nm. Initially the disks were saturated along [10] and [11] ($H_{\mathrm{ext}}$) followed by relaxing in a external applied field of 0 mT. The lighter regions along the disk perimeters illustrate the strength of the interaction of the disks. The letters represent the magnetic texture and arrows the orientation of the mesospin moments.
    			}
	\end{figure}

    Spatial maps of the amplitude of the demagnetizing field for mesospins with a diameter of 350 nm are shown in Fig. \ref{fig:fig7}. The system was initially saturated along the [10] and [11] axis (illustrated by $H_{\mathrm{ext}}$), followed by reducing the field to zero. This results in different magnetic textures stabilizing, indicated by the letters at the center of the disks, with the arrows specifying the rotation of $|\langle \mathbf{m} \rangle |$. The lighter regions in these magnetostatic field maps represent a bond between the islands. 
    
    The results displayed in Fig. ~\ref{fig:fig7}a arise from dressing the mesospins with a field along the horizontal axis. The overall magnetic texture consists of O-states, with a mesospin moment along the direction of the externally applied field. As seen in the figure, the attractive interaction is confined to the nearest neighboring mesospins along [10]. Comparing $|\langle \mathbf{m} \rangle |$ for the O-state in Table ~\ref{table1} and $M_{\mathrm{m}}$ in Fig. ~\ref{fig:fig5}a, reveals a lower net moment in the experiments as compared to the fully dressed mesospins. Capturing these changes therefore requires a rotation of the mesospins and/or a presence of other states, resulting in a reduced net moment along the [10] direction. In Fig.~\ref{fig:fig7}b the external field has been applied along [11], resulting in islands with both C- and S-states with rotated mesospin moments. Furthermore, for this applied field direction, mesospins can link up to four neighbors. This can be seen for the case of the center island, having an S texture and strong amplitudes of the magnetostatic field towards its four nearest neighbors. The possibility of mixing mesospin textures of more than just one type, having further considerable transverse magnetization components, explains the consistently lower values of $M_{\mathrm{m}}$ shown in Fig. \ref{fig:fig5}a and the [11] direction compared to the [10] direction.
    
\section{\label{sec:conclusion}Conclusions}
	
    We provide experimental evidences of a configurational dependence in the energy of mesospins forming a square lattice. An emerging anisotropy is obtained, in line with previous findings \cite{Mathieu1997, Natali2002, Ciuciulkaite2019, Kakazei2006, Xiaobin2002}. We demonstrate the effects of the anisotropy in the interaction energy by determining the directional dependence of the thermally induced transition from dressed collinear states to vortex states of the elements. The key component required for rationalizing the results is an interplay between the interaction of the mesospins and their magnetic texture: When a mesospin and its neighbors have a net magnetization along [10] in the square lattice, the interaction can be viewed as being one dimensional with the interaction restricted to two neighbors along the [10] axis. When the net moment is along the [11], the elements can have interactions with four neighbors. These differences are caused by self-induced modification of the inner magnetic texture, resulting in changes of the interaction potential of the mesospins. The spontaneous interactions of elements below the critical limit are therefore expected to bias the [11] configuration, while a mixed order with different range of correlations is expected to emerge at finite temperatures. 
	
\section*{\label{sec:acknowledgement}Acknowledgements}
B.H. and V.K. acknowledge financial support from the National Research Council (VR) (Project No. 2019-03581 and {2019-05379}). We acknowledge Myfab Uppsala for providing facilities and experimental support. Myfab is funded by the Swedish Research Council (2019-00207) as a national research infrastructure.

\section*{Author Declarations}

\subsection*{Data availability}
The data that support the findings are available from the corresponding authors upon reasonable request.

\subsection*{Conflict of Interest}
The authors have no conflicts to disclose.

%	\bibliography{dots}	
%	\bibliographystyle{apsrev4-1}

	
\end{document}